
\input phyzzx
\def\bal{$\alpha$\kern-.7em\hbox{$\alpha$}}

\sequentialequations
\Pubnum{ }
\titlepage
\title{\centerline{A Mean Field Theory for the Quantum Hall Liquid. II}
       \nextline
      \centerline{--- The Vortex Solution---}}
\author{\rm Kenzo I{\tenrm SHIKAWA} and Nobuki M{\tenrm AEDA}}

\address{Department of Physics, Hokkaido University, Sapporo 060 JAPAN}
\abstract

In the Fractional Quantum Hall state,
we introduce a bi-local mean field and get vortex
mean field solutions.
Rotational invariance is imposed and
the solution is constructed by means of numerical self-consistent method.
It is shown that vortex has a
fractional charge, a fractional angular momentum and a
magnetic field dependent energy.
In $\nu=1/3$ state, we get finite energy gap at $B=10,15,20[T]$.
We find that the gap vanishes at $B=5.5[T]$ and becomes negative below it.
The uniform mean field becomes unstable toward vortex pair production
below $B=5.5[T]$.
\endpage

\doublespace

\chapter{Introduction}
In previous papers\Ref\a{K.Ishikawa,Prog.Theor.Phys,
Suppl.No.107(1992)167;Vol.88(1992)881. },
we introduced a bi-local mean field
in two dimensional electron system in perpendicular magnetic field and
discussed a uniform ground state.
In this paper, we discuss a vortex solution
which is a non-linear and topological excitation of the system.
Experiment of the fractional quantum Hall effect shows that
the excited state has the energy gap
and the ground state becomes the incompressible liquid.
An interesting liquid state that has these properties
was proposed by Laughlin\Ref\Lau{
R.B.Laughlin,Phys.Rev.Lett.{\bf50}(1983)1395.}.
In Laughlin wave function the excited state includes quasiholes and
quasiparticles. They give a finite energy gap of the system.
Experimentally it is mesured as activation energy.
The other interesting properties of Laughlin wave function are that
quasiholes and quasiparticles have fractional charges and
obey fractional statistics.
Theoretical calculations based on Laughlin wave function
\REF\rj{R.Morf and B.I.Halperin,
Phys.Rev.B33(1986)2221}
\REF\rc{S.M.Girvin,A.H.MacDonald and P.M.Platzman,
Phys.Rev.Lett.{\bf 54}(1985)581,
\nextline Phys.Rev.B33(1986)2481}
\refmark{\Lau-\rc} and
on exact diagonalization of the system with finite numbers
of electrons
\REF\rb{D.Yoshioka,B.I.Halperin and P.A.Lee,Phys.Rev.Lett.
{\bf 50}(1983)1219}
\REF\ra{F.D.M.Haldane and E.H.Rezayi,Phys.Rev.Lett.
{\bf 54}(1985)237}
\REF\rd{W.P.Su,Phys.Rev.B32(1985)2617}
\REF\re{G.Fano,F.Oryolani and E.Colombo,Phys.Rev.B34(1986)2670}
\REF\rf{A.H.MacDonald and G.C.Aers,Phys.Rev.B34(1986)2906}
\REF\rn{D.Yoshioka,J.Phys.Soc.Jpn,{\bf 55}(1986)885}
\refmark{\rb-\rn}
 predict that activation energy is
roughly proportional to $\sqrt B$.
Experimentally activation energy
is smaller than the theoretical values and
does not follow $\sqrt B$ dependence.
It vanishes below $B\sim5[T]$ and follows roughly
liner dependence of $B$ above it\Ref\bc{
{\it The Quantum Hall Effect} ed. R.E.Prange and S.M.Girvin
(Springer, New York, 1990) and references therein.
}.
Laughlin wave function is a trial function, then
it may not be good enough for more systematic investigation.
For this porpose, we start from a microscopic many-body
Hamiltonian and apply a mean field theory to it.
We study the vortex solutions in our theory which
may correspond to Laughlin's quasihole and
quasiparticle.

It is well known that Abelian Higgs theory
has the vortex solution which is described
by the classical solution of bosonic fields.
We construct its counterpart in two dimensional electron system
in perpendicular magnetic field based on the bi-local mean field theory.
The elementary field is fermionic in this system,
which does not have classical expectation value.
We treat two point function $\VEV{\psi^\dagger(y)\psi(x)}$,
instead of one point function of bosonic variables,
as the mean field and find the vortex solution.
This mean field is bi-local and fluctuations of the mean field
includes a dynamical gauge field
and a density fluctuation field.
The mean field satisfies a self-consistency equation, which becomes
to a non-linear Schr\"odinger equation for these fields.
The solutions are found numerically. Vortices have similar
properties as Laughlin's quasihole and quasiparticle by having
fractional charges and fractional angular momenta.
The $B$ dependence of activation energy is similar to experimental
data rather than Laughlin's one and is smaller than both
of Laughlin's one and experimental one.
We find that energy gap vanishes at $B=5.5[T]$
in agreement with experimental data.

The present paper is organized in the following way.
In section 2, we study the non uniform density solution of
the self-consistency equation.
In section 3, we calculate the vortex solution at $\nu=$1/3
by the iteration method .
The shape, the charge and the angular momentun of
the vortex are presented. We calculate magnetic field dependence
of gap energy and find that there exists a threshold of
the magnetic field
for our uniform ground state to be stable.
In section 4, summary and discussion are given.

\chapter{Mean Field and Vortex Solution}
The Hamiltonian of two dimensional non-relativistic
electrons in the
perpendicular constant magnetic field B is
$$
H=\int d^2x\psi^\dagger(x){({\bf P}+e{\bf A})^2\over2m}\psi(x)
+{1\over2}\int d^2xd^2y\psi^\dagger(x)\psi^\dagger(y)V(x-y)
\psi(y)\psi(x).
\eqno\eq
$$
where
$
P_i=-i\partder{}{x^i},
\partial_1A_2-\partial_2A_1=B,\
V(x-y)={{\textstyle e^2}\over\vert x-y\vert} .
$
We neglect inessential background terms\refmark{\a}.
In the functional integral formalism,
the partition function is written as
$$
Z=Tr(e^{-\beta H})
=\int D\psi^\dagger D\psi e^{-\int^\beta_0 d\tau[\int d^2x\psi^\dagger
\partial_\tau\psi+H]}.
\eqno\eq
$$
We introduce the bi-local
auxiliary field $U(x,y)$, and $Z$ is written as
$$
Z=\int DU D\psi^\dagger D\psi
e^{-\int^\beta_0 d\tau[\int d^2x\psi^\dagger
\partial_\tau\psi+H^U]},
\eqno\eq
$$
$$
\eqalign{
H^U&=\int d^2x\psi^\dagger(x){({\bf P}+e{\bf A})^2\over2m}\psi(x)\cr
&+{1\over2}\int d^2xd^2y V(x-y)[\vert U(x,y)\vert^2-
U(x,y)\psi^\dagger(x)\psi(y)-
U^*(x,y)\psi^\dagger(y)\psi(x)].
}
\eqn\H
$$
We take $\beta\rightarrow\infty$.
$U(x,y)$ is stationary at the mean field $U_0(x,y)$
which is given by
$$
U_0(x,y)=\VEV{\psi^\dagger(y)\psi(x)}.
\eqn\bo
$$
We assume the following ansatz,
$$
U_0(x,y)=U_0\rho(x,y)e^{-\gamma(x-y)^2}\exp[i\int\nolimits
^y\nolimits_x \alpha_id\xi^i],
\eqn\U
$$
where $\rho$ is real symmetric function of $x$ and $y$, and
$U_0={\nu/\pi R^2_0},
\gamma={1/2R^2_0},R_0=\sqrt{2/eB}$. Line integral
is along a straight line between $x$ and $y$.
We assume, further, $\lim\limits_{x\rightarrow\infty,y\neq0}\rho(x,y)=1$.
These parameters are fixed in order that $U(x,y)$ may
be coincident with a uniform self-consistent solution\refmark{\a}
at infinity.
The self-consistent solution for uniform ground state is
obtained by replacing $\rho$ by 1 and set $\alpha_i=eA_i$
at Eq.\U,
$$
U_0^{(g.s.)}(x,y)=U_0e^{-\gamma(x-y)^2}\exp[i\int\nolimits
^y\nolimits_x eA_i d\xi^i].
\eqn\UG
$$

By substituting Eq.\U\ into Eq.\H, we get the mean field Hamiltonian,
$$
\eqalign{
H_0=&\int
d^2x \lim_{{\bar x}\rightarrow x}
\psi^\dagger(x)[{({\bf P}+e{\bf A})^2\over2m}-
F(({\bf P}+\hbox{\bal})^2)\rho({\bar x},x)]
\psi(x)\cr
 &+{1\over 2}\int d^2xd^2y V(x-y)\vert U_0(x,y)\vert^2,\cr
F(p^2)&=e^2U_0{\pi^{3/2}\over\gamma^{1/2}}e^{-p^2/8\gamma}
I_0(p^2/8\gamma). \cr
}
\eqn\F
$$
Where $I_0$ is modified Bessel function.
{}From Eq.\F, single-body Hamiltonian reads
$$
H_s =[{({\bf P}+e{\bf A})^2\over2m}-
F(({\bf P}+\hbox{\bal})^2)\rho({\bar x},x)]_{\bar x\rightarrow x}.
\eqn\s
$$
The second term is mean field's contribution through a
Coulomb interaction.
The system is invariant under a gauge transformation,
$$
\eqalign{
\psi &\rightarrow e^{i\Lambda(x)}\psi, \cr
A_i &\rightarrow A_i -{1\over e}\partial_i \Lambda(x), \cr
\alpha_i &\rightarrow \alpha_i-\partial_i \Lambda(x).}
\eqno\eq
$$
The electron field is expanded by $H_s$'s eigenfunctions.
$$
\eqalign{
\psi(x)&=\sum_{l,n}u_{l,n}(x)a_{l,n},\cr
 H_s u_{l,n}(x)=E_{l,n}&u_{l,n}(x),\ \{a^\dagger_{l,n},a_{l',n'}\}
=\delta_{l,l'}\delta_{n,n'},\cr
E_{l,0}<&E_{l,1}<E_{l,2}<\cdots
}
\eqno\eq
$$
where $l$ is angular momentum quantum number
and $a^\dagger_{l,n}$,$\ a_{l',n'}$ are anti-commuting
creation and annihilation operators.
The N-electron state is made as
$$
\ket{\psi}=\sum_{l_1,l_2\cdots l_N=0}^M F_{l_1,\cdots l_N}
a^{\dagger}_{l_1,0}a^{\dagger}_{l_2,0}\cdots a^{\dagger}_{l_N,0}
\ket{0}.
\eqno\eq
$$
We restrict the state within the lowest energy level for every $l$
and assume that
$\ket{\psi}$ is the superposition of same angular
momentum states. Then
$$
\sum_{l_2\cdots l_N=0}^M F^*_{n,l_2,\cdots l_N}F_{m,l_2,\cdots l_N}
=\nu_n\delta_{n,m},\quad 0\leq \nu_n \leq 1,
\eqno\eq
$$
and two point function is
$$
\bra{\psi}\psi^\dagger(y)\psi(x)\ket{\psi}
=\sum_{l=0}^M \nu_l u^*_{l,0}(y)u_{l,0}(x).
\eqn\ps
$$
{}From Eq.\bo, Eq.\U\ and Eq.\ps, we get a self-consistency condition
$$
U_0\rho(x,y)e^{-\gamma(x-y)^2}\exp[i\int\nolimits
^y\nolimits_x \alpha_id\xi^i]=\sum_{l=0}^M
\nu_l u^*_{l,0}(y)u_{l,0}(x).
\eqn\self
$$
Uniform density solution Eq.\UG\ satisfies Eq.\self\
for $\nu_l=\nu$,
$$
\sum_{l_2\cdots l_N=0}^M F^{(g.s.)*}
_{n,l_2,\cdots l_N}F^{(g.s.)}_{m,l_2,\cdots l_N}
=\nu\delta_{n,m}.
\eqno\eq
$$
In order that the state may be coincident with a uniform density
state of filling $\nu$ at infinity,
$\nu_l$ must approach to the value $\nu$
for large $l$.

Let us construct the vortex solution in this formalism.

In the Abelian Higgs theory, the vortex solution
\ref{A.Abrikosov,JETP(Sov.Phys.){\bf5}(1957)1174.\nextline
H.Nielsen and P.Olesen,Nucl.Phys.B61(1973)45.}
takes the form $\VEV{\phi(r,\theta)}\sim v(r)e^{in\theta}$, $n$
is integer.
$n$ is winding number of a mapping :
$S^1\rightarrow U(1)$.
At infinity $v$ goes to its vacuum expectation value
and at $r=0$, $v$ is zero.
For the finiteness of the energy, the kinetic term
$\vert{\bf\partial}\phi+ie{\bf A}\vert^2$
must vanish at infinity. Then gauge potential must take the form
$A_{\theta}\sim-n/e$. Magnetic field localizes around the vortex.

By similar considerations,
it is natural to impose the following ansatz in our case,
$$
\eqalign{
\alpha_i(x)&=eA_i(x)+n_i(x),\quad
{\bf n}(x)={n(r)\over r^2}(-x_2,x_1),\cr
&{\bf A}(x)={B\over2}(-x_2,x_1),\cr
}
\eqn\a
$$
and boundary conditions are
$$
\eqalign{
&n(0)=n\ ;\ {\rm integer},\cr
&\lim_{r\rightarrow\infty}n(r)=0,\cr
&\rho(x,0)=\rho(0,x)=0,\ \rho(x,x)=\rho(r).}
\eqn\bou
$$
Dynamical gauge field $\alpha_i$ has singularity at origin
for $n(0)\neq 0$.
But this singularity is harmless in Eq.\F,
since $\rho$ vanishes at origin.
Therefore eigenfunction $u_l$ behaves regularly as
$r^l$ near the origin.
By the singular gauge transformation,
$
\psi \rightarrow e^{in\theta},\
\alpha_i \rightarrow \alpha_i-n\partial_i\theta,
$
 this singularity is removed.
Then $n$ is regarded as winding number in the same way as
Abelian Higgs theory.

In Eq.\bou, $n$ is integer for the reason that $U(x,y)$ must be
continuous function of $x$ and $y$.
Fig.1 shows a line integral along a infinitesimal circle $c$
around origin, which is
$$
\oint_c \alpha_i d\xi^i=2\pi n(0),
\eqno\eq
$$
generates discontinuity
when the integration path cross the origin.
Therefore, $n(0)=n$ must be integer for continuity.
\FIG\c{The path of line integal in $U_0$
is drawn. \nextline (a)
Fixing a point $x$, a point $y$ is moved around origin.
\nextline (b)
Infinitesimal circle $c$ around origin gives a discontinuity $2\pi n$.}

For the finite energy, $\alpha_i$ has to coincide with
the external field $A_i$ at infinity. See Appendix~A
where we present exact soluble example
which shows that if $n(\infty)\neq 0$ or $\alpha_i(\infty)
\neq eA_i$ then a vortex excitation energy diverges.
Moreover the rotational invariance implies the above form
of $\alpha_i$. See Appendix B.

Momentum and angular momentum density operator are
defined by
$$
\eqalign{
P_i(x)&={1\over2}\{\psi^\dagger(x)(-i\partial_i+eA_i(x))\psi(x)+
               [(i\partial_i+eA_i(x))\psi^\dagger(x)]\psi(x)\},\cr
L(x)&=\epsilon_{ij}x_i P_j(x).}
\eqno\eq
$$
Using Eq.\U\ and Eq.\a, the expectation values are given by
$$
\eqalign{
\VEV{P_i(x)}&=-U_0\rho(r)n_i(x),\cr
\VEV{L(x)}&=-U_0\rho(r)n(r)
,}
\eqn\V
$$
We can see from Eq.\V\ that the current is in proportion to $\rho$
and a difference between $A_i$ and $\alpha_i$.
This corresponds to London equation in the theory
of superconductivity.
For the $n_i$ given by Eq.\a,
the current is rotating around the vortex.

\chapter{Numerical Solution of Vortex}
In Eq.\F, $F(p^2)$ is difficult to calculate numerically.
We expand $F(p^2)$ around $p^2=eB$ which corresponds to energy
of lowest Landau level
and approximate as
$$
\eqalign{
F(p^2)&=F(eB)+F'(eB)(p^2-eB)
=F_0-{p^2\over 2m'},\cr
 F_0&=e^2 U_0 {\pi^{3/2}\over{\gamma^{1/2}}}
\exp(-{1\over2})[3I_0({1\over2})-I_1({1\over2})]/2,\cr
{1\over 2m'}&={e^2\over 8} U_0 ({\pi\over\gamma})^{3/2}
\exp(-{1\over2})[I_0({1\over2})-I_1({1\over2})],
}
\eqn\Fp
$$
where $I_n$ is modefied Bessel function.
Then we have to solve the following non-linear
Schr\"odinger equation,
$$
\eqalign{
[{({\bf P}+e{\bf A})^2\over2m}&+{({\bf P}+\hbox{\bal})^2\over2m'}
\rho(\bar x,x)+F_0(1-\rho(x))]_{\bar x\rightarrow x}
u_l(x)=E_l u_l(x),\cr
u_l(x)&=v(r)e^{-il\theta}\ ;\ l\ {\rm is\ integer},\cr
}
\eqn\non
$$
where $\alpha_i$ and $\rho$ are given by self-consistency condition
Eq.\self.
We add a constant $F_0$ to energy for simplicity.
The $\rho$ looks like a scalar potential.
We make a quasihole state by removing $l=0$ state
from uniform density state as
$$
\ket{hole}=\sum_{l_1,l_2,\cdots,l_N=0}^M
F_{l_1,\cdots,l_N}^{(g.s.)}
a^{\dagger}_{l_1+1,0}a^{\dagger}_{l_2+1,0}\cdots a^{\dagger}_{l_N+1,0}
\ket{0}.
\eqn\hs
$$

Hereafter lengths are defined in units of $R_0$.
We fix the form of $n(r)$ for $n(0)=1$ as
$$
n_{+}(r)={r^2\over e^{r^2}-1}.
\eqn\n
$$
This form is implied by removing $l=0$ state from
the lowest Landau level.
That is,
$$
\eqalign{
\VEV{\psi^\dagger(y)\psi(x)}_{{\rm lowest\ Landau\ level},l\ne0}
&={\nu\over\pi}[\sum^\infty_{l=1}{(z_y z_x^*)^l\over l!}
e^{-{1\over2}(\vert z_x\vert^2+\vert z_y\vert^2)}]\cr
&={\nu\over\pi}(1-e^{-z_y z_x^*})
e^{-{1\over2}\vert z_x-z_y\vert^2+i\int^y_x eA_i d\xi^i},
}
\eqno\eq
$$
where $z_x=x_1+ix_2$.
The phase part of
$1-e^{-z_y z_x^*}$ is coincident with Eq.\n\
for $\vert z_x-z_y\vert\ll1$. $\rho$ reads
$$
\rho_0(x,y)=\sqrt{1+e^{-2r_x r_y\cos\theta}-2\cos(r_x r_y\sin\theta)
e^{-r_x r_y\cos\theta}}.
\eqno\eq
$$
This has properties as
$\rho_0(x,y)=\bar\rho_0(\sqrt{r_x r_y},\theta);\bar\rho_0(r,0)=\rho_0(r)$.
We impose these to the $\rho$ and calculate $\rho$ by
numerical self-consistent method.
We find that ansatz Eq.\n actually satisfies self-consistency condition
Eq.\self.

By using Eq.\n, we get eigenvalue $E_l$ for $\nu={1\over3}$,
\foot
{
We have performed the numerical calculation for $l=0,1,\cdots,30$
and interpolate for large $l$ to the uniform solutions.
}
and find that $E_0$ is bigger than the other $E$'s.
Then we remove the $l=0$ state and calculate the $\rho(r)$ by Eq.\self\
and using it we calculate $E_l$ again.
By iterating this procedure, we get the self-consistent solution.
Final results for eigenvalues are listed at
Table I for several value of $B$ and see Fig.2.
The $\rho(r)$ is shown in Fig.3.

\FIG\d{Eigenvalues at $B=$15$[T]$, $\nu=1/3$. \nextline
$\bullet$ represents eigenvalue for quasihole and
$\circ$ represents eigenvalue for quasiparticle. }

\FIG\a{$\rho(r)$ for quasihole at $B=$15$[T]$, $\nu=1/3$. }

Asymptotically $\rho(r)\sim 1-e^{-r^2}$ at infinity.
This behavior is different from the vortex in Higgs field.
That is
$v(r)\sim v(1-e^{-cr})$.
And also, dynamical gauge field behaves as
$n(r)\sim e^{-r^2}$ at infinity.
These properties are characteristic of the system in magnetic
field.

Apparently, the charge decreases by $\nu e$
as compared with the uniform density state.
Then this solution corresponds to quasihole
which has fractional charge $Q_+=-\nu e$. That is
$$
Q_+=eU_0\int d^2x (\rho(r)-1)=-\nu e.
\eqno\eq
$$
This relation is exact.

Angular momentum of vortex is given by Eq.\V\ as
$$
\VEV{L}=\int d^2x \VEV{L(x)}=-U_0\int d^2x \rho(r)n(r).
\eqn\L
$$
This value is calculated numerically
at $\nu=1/3$ as $-0.989/3$, $-0.991/3$, $-0.993/3$, $-0.994/3$ at $B=$
5, 10, 15, 20$[T]$
respectively.
Then the angular momentum of vortex for quasihole is
approximately $-1/3$.

In order to get the gap energy, we need a
quasiparticle solution which has a opposite charge to quasihole.
We make a quasiparticle state by removing $l=0$ state from
uniform density state and filling $l=1$ state fully as
$$
\ket{particle}=a^{\dagger}_{1,0}\sum_{l_1,l_2,\cdots,l_N=0}^M
F_{l_1,\cdots,l_N}^{(g.s.)}
a^{\dagger}_{l_1+2,0}a^{\dagger}_{l_2+2,0}\cdots a^{\dagger}_{l_N+2,0}
\ket{0}.
\eqn\ps
$$
We fix $n(r)$ for $n(0)=1$ as
$$
n_-(r)={r^2(3-2r^2)\over{e^{r^2}+2r^2-1}}.
\eqn\nm
$$
This form is implied by removing $l=0$ state from
the lowest Landau level and filling $l=1$ state fully.
In Eq.\nm, $n_-$ expresses a current which is rotating
around the vortex and the direction is reversed at
$r=\sqrt{3/2}$.
We calculate the eigenvalues $E_l$ for $\nu=1/3$ and find that
$E_0$ is bigger and $E_1$ is smaller than the others.
Then, it is natural to
remove the $l=0$ state and fully fill the $l=1$ state,
\ie , $\nu_0=0,\ \nu_1=1,\  \nu_l=1/3 ; l\ge2$. This construction
means that $Q_-=(-1/3+2/3)e=(1/3)e$.
This state corresponds to quasiparticle
which has fractional charge $Q_-=\nu e$ for $\nu=1/3$.
We perform the same iteration procedure as quasihole.
We find that ansatz Eq.\nm actually satisfies self-consistency condition
Eq.\self.
Final results for eigenvalues are listed at
Table II for several value of $B$ and see also Fig.2.
The $\rho(r)$ is shown in Fig.4.

\FIG\b{$\rho(r)$ for quasiparticle at $B=$15$[T]$, $\nu=1/3$. }

By Eq.\L, the angular momentum is calculated numerically as
0.965/3, 0.978/3, 0.983/3, 0.986/3 at $B=$
5, 10, 15, 20$[T]$
respectively.
Then the angular momentum of vortex for quasiparticle is
approximately 1/3.

The gap energy $\Delta$ is half of the pair excitation energy of
a quasihole and a quasiparticle which are separated infinitely.
That is
$$
\Delta=(E_{hole}+E_{particle})/2.
\eqno\eq
$$
Experimentally $\Delta$ is measured as activation energy.
It is determined from temperature dependence of
diagonal resisitivity as $\rho_{xx}\propto e^{-\Delta/T}$.
Each energy is calculated as
$$
E_{vortex}=\sum_l [\nu_l E_l-\nu E_l^{(g.s.)}]+
{1\over 2}\int d^2x d^2y V(x-y)[\vert U_0(x,y)\vert^2-
\vert U_0^{(g.s.)}(x,y)\vert^2],
\eqno\eq
$$
where $E_l^{(g.s.)}$ is energy eigenvalue of uniform ground state
and $U_0^{(g.s.)}$ is given by Eq.\UG.
For several
values of $B$, gap energies are listed in Table III.
Finite energy gap exists at $B=10,\ 15,\ 20[T]$ and
the gap becomes negative at $B=5[T]$.
See Fig.5.
\FIG\c{The gap energy vanishes at $B=5.5[T]$ and increases monotonically
above it.}
The calculations based on Laughlin wave function\refmark{\Lau-\rc}
and on exact diagonalization of the finite system
\refmark{\rb-\rn} show
$\Delta(B)\propto\sqrt B$.
Our result shows $\Delta$ increases linerly and
vanishes at $B=5.5[T]$.
The $B$ dependence of gap energy is similar to the experiments
\refmark{\bc}.
Experiments show that gap energy increases monotonically with $B$ and
there exists threshold of finite gap at $B\sim5[T]$.
The theoretical calculations including the effect of disorder
yeild threshold effect
\REF\rh{A.H.MacDonald,K.L.Liu,S.M.Girvin and P.M.Platzman, \nextline
Phys.Rev.B33(1986)4014}
\REF\ri{A.Gold,Phys.Rev.B33(1986)5959}
\refmark{\rh,\ri}.
However, our result means that threshold effect
remains without the effect of disorder.

Note that the quasiparticle solution is constructed above
not for $n(0)=-1$ but for $n(0)=1$. The state contains
eigenfunction $u_l$'s only for $l\ge0$
and the phase dependence is $e^{il\theta}$, then $n(0)=-1$,\ie,
$l=-1$, can't be induced.
The energy for $l\le-1$ belongs to the higher Landau level.
In other words, the solution for $n(0)=-1$ can not be constructed
self-consistently in lowest Landau level.

\chapter{Summary and Discussion}
We construct the vortex solution in the bi-local mean field
theory by numerical method.
By dynamical gauge field which has singularity at
core of vortex, energy level
becomes non-degenerate for small $l$ and charge density
decreases or increases near the vortex.
Their charges are exactly $\pm e/3$ at $\nu=1/3$
and both of them have winding number $n=1$.
Their angular momenta are approximately $\mp 1/3$.
Then statistics of the vortex is supposed to be fractional.
\ref{F.Wilczek, Phys.Rev.Lett.{\bf48},(1982),1144.}

We get finite energy gap at $B>5.5[T]$ and $\nu=$1/3,
then the ground state is incompressible.
Below $B=$5.5$[T]$, the energy gap takes negative value.
This result means that the uniform mean field solution
becomes unstable and activation energy vanishes at $B<5.5[T]$.
Thus we have obtained magnetic field threshold effect
without disorder for the first time.
Experiment of mobility dependence of threshold effect
will discriminate our calculation from other disorder theory.
\Ref\rg{R.L.Willet,H.L.Stormer,D.C.Tsui,A.C.Gossard \nextline
and J.H.English, Phys.Rev.B37(1988)8476}
Other theoretical calculations without disorder did not show
threshold effect.
They include only lowest Landau level and next Landau level at most.
On the other hand our method includes Landau levels
higher than next Landau level.
Especially at low magnetic fields,
higher Landau levels become important.
Furthermore, since vortex solutions are highly non-linear objects,
the projection on the lowest Landau level may not be good approximation.
Our results suggest that
the threshold effect may be related to higher Landau levels.

For large $B$, our energy gap is much smaller than
the value of experiments.
However experimental data suggest the existence of second
activation energy at lower temperature.
\REF\rk{G.S.Boebinger, A.M.Chang, H.L.Stormer and D.C.Tsui, \nextline
Phys.Rev.Lett.{\bf 55}(1985)1606}
\REF\rmm{J.Wakabayashi, S.Kawaji, J.Yoshino and H.Sakaki, \nextline
J.Phys.Soc.Jpn.{\bf 55}(1986)1319}
\REF\rl{G.S.Boebinger, H.L.Stormer, D.C.Tsui, A.M.Chang,
J.C.M.Hwang, \nextline A.Y.Cho, C.W.Tu and G.Weimann,
Phys.Rev.Lett.{\bf 36}(1987)7919}
\refmark{\rk-\rl}
Our results are consistent with the second activation energy.
But hopping conduction
\REF\rp{Y.Ono,J.Phys.Soc.Jpn.{\bf 51}(1982)237}
\REF\rq{N.F.Mott and E.A.Davis, {\it Electronic Properties in
Non-Crystalline Materials}, 2nd ed. (Clarendon, Oxford, 1979)}
\refmark{\rp,\rq}
 is another candidate for lower temperature
behavior of $\rho_{xx}$.
Thus we wish more experiments to be done at lower temperature.

We construct a excited state by Eq.\hs\ and Eq.\ps.
In addition to these state, our formalism must include
other excited states.
Then more than one gap energies may exist.

The following improvements for our method may be possible.
The first is to take higher derivative terms in Eq.\Fp.
Higher derivative terms may be irrelevant in long distance physics.
However these terms affect Eq.\non\ and energy eigenvalues may
change. The second is to calculate $n(r)$ by numerical iteration
method self-consistently.

\ack
The present work is partially supported by a Grant-in-Aid for
general Scientific Research (03640256), and the Grant-in-Aid for
Scientific Research on Priority Area (04231101).

\Appendix {A}
We consider here the soluble eigenvalue problem of
following Hamiltonian,
$$
H={({\bf P}+{\bf A})^2\over2m}+{({\bf P}+{\bf A'}+{\bf n})^2\over2m'},
\eqno\eq
$$
$$
{\bf A}={B\over2}(-x_2,x_1),\ {\bf A'}={B'\over2}(-x_2,x_1),\
{\bf n}={n\over r^2}(-x_2,x_1),
\eqno\eq
$$
where $n$ is constant value.
The eigenfunction is
$$
\eqalign{
\psi_{N,l}=&Const\cdot r^{ \tilde l} e^{-il\theta-({\tilde B}/4)r^2}
u_{N,l}(r),\ l=0,\pm1,\pm2,\cdots,\cr\
&u_{N,l}(r)=L^{(\tilde l)}_N({\tilde B\over2}r^2),\cr
{\tilde l}=&[M(l^2/m+(l-n)^2/m')]^{1/2},\ 1/M=1/m+1/m',\cr
{\tilde B}=&[M(B^2/m+B'^2/m')]^{1/2},\cr
}
\eqno\eq
$$
where $L_N^{(\tilde l)}$ is associated Laguerre function.
The eigenvalue is
$$
E_{N,l}={{\tilde B}\over M}(N+{1\over2})+
[nB'/m'-(B/m+B'/M')l+{\tilde l}{\tilde B}/M]/2.
\eqno\eq
$$
(a) $B\ne B'$ case
$$
E_{N,l}-E_{N,l}^{n=0}{\atop\widetilde{l\rightarrow\infty}}
C\cdot l,
\eqno\eq
$$
$\qquad
C=[\sqrt{(B/m+B'/m')^2+(B-B')^2/mm'}
-(B/m+B'/m')]/2>0.
$

Energy difference diverges in proportion to $l$.

\noindent
(b) $B=B'$ case
$$
E_{N,l}-E_{N,l}^{n=0}{\atop\widetilde{l\rightarrow\infty}}
{\vert B\vert n^2\over 4(m+m')}\cdot {1\over l}
\eqno\eq
$$
Energy difference does not diverge, but the sum $\sum{1\over l}$
diverges.

\Appendix {B}
In the uniform magnetic field, the system has translational and
rotational invariance. But, in the quantum theory, magnetic field affects
the system in the form of gauge potential. Therefore the symmetries seem
to be lost in the system. Since the system also has gauge symmetry,
translational and rotational invariance remain in the quantum theory
\ref{S.Randjbar-Daemi,A.Salam and J.Strathdee,Nucl.Phys.B340
(1990)403,\nextline see Appendix A therein.}.

At first, we consider the translation
$x_i'=x_i+\epsilon_i$. Vector potential is transformed as
$$
A_i\rightarrow A'_i(x')=A_i(x)+\partial_i\Lambda(x,\epsilon)
\eqno\eq
$$
Translational invariance means $A'_i(x)=A_i(x)$, then
$$
\partial_i\Lambda(x,\epsilon)=A_i(x+\epsilon)-A_i(x).
\eqno\eq
$$
If the magnetic field
$B(x)=B(x+\epsilon)$ then $\Lambda$ exsists and given by
$$
\Lambda(x,\epsilon)=\int\nolimits^x [A_i(\xi+\epsilon)-A_i(\xi)]d\xi^i
\eqn\La
$$

Our Hamiltonian is made from two gauge field, that is,
$(P+A)^2/2+F(P+\alpha)$. Hence the $\Lambda$ made by $A_i$ and
$\Lambda'$ made by $\alpha_i$ must be same. From Eq.\La, this means that
translational invariance only allows the difference of a
constant vector between $A_i$ and $\alpha_i$.
In fact, we know that when $A_i=\alpha_i$, uniform self-consistent solution
exists.

Next we consider the rotation,
$x'_1=x_1\cos\theta+x_2\sin\theta,\ x'_2=-x_1\sin\theta+x_2\cos\theta$.
Vector potential is transformed as
$$
\cases{
&$A_1\rightarrow A'_1(x')=A_1(x)\cos\theta+A_2\sin\theta
+\partial_1\Lambda(x,\theta)\cos\theta+
\partial_2\Lambda(x,\theta)\sin\theta$\cr
&$A_2\rightarrow A'_2(x')=-A_1(x)\sin\theta+A_2\cos\theta
-\partial_1\Lambda(x,\theta)\sin\theta+
\partial_2\Lambda(x,\theta)\cos\theta$\cr
}
\eqno\eq
$$
If $B(x)=B(r)$ then $\Lambda$ exsists and given by
$$
\eqalign{
\Lambda(x,\epsilon)&=\int\nolimits^x
[\cos\theta A_1(\xi')-\sin\theta A_2(\xi')-A_1(\xi)]d\xi^1\cr
&+[\sin\theta A_1(\xi')+\cos\theta A_2(\xi')-A_2(\xi)]d\xi^2
}
\eqn\la
$$

In order that the $\Lambda$ made by $A_i$
and $\Lambda'$ made by $\alpha_i$ may be same, their difference must
be the following form
$$
{\bf A}(x)-\hbox{\bal}(x)=f(r)(x_1,x_2)+g(r)(-x_2,x_1).
\eqno\eq
$$
One can easily see that Eq.(2.17) satisfies this condition.

\refout
\endpage
\figout
\end

\documentstyle[12pt]{article}
\begin{document}
\pagestyle{empty}

Table.1

\

\begin{tabular}{|c|c|c|c|c|}\hline
l &$E_l(B= 5)$ &$E_l(B=10)$ &$E_l(B=15)$ &$E_l(B=20)$ \\ \hline
0 &0.733     &0.680     &0.653     &0.636 \\
1 &0.593     &0.571     &0.560     &0.553 \\
2 &0.533     &0.525     &0.522     &0.519 \\
3 &0.508     &0.507     &0.506     &0.505 \\
4 &0.500     &0.500     &0.500     &0.500 \\
5 &0.498     &0.499     &0.499     &0.499 \\
6 &0.498     &0.499     &0.499     &0.499 \\
7 &0.499     &0.499     &0.499     &0.499 \\
8 &0.499     &0.499     &0.499     &0.500 \\ \hline
\end{tabular}

\

\

Table.2

\

\begin{tabular}{|c|c|c|c|c|}\hline
l &$E_l(B= 5)$ &$E_l(B=10)$ &$E_l(B=15)$ &$E_l(B=20)$ \\ \hline
0 &0.565     &0.557     &0.551     &0.547 \\
1 &0.372     &0.404     &0.420     &0.430 \\
2 &0.378     &0.408     &0.423     &0.432 \\
3 &0.426     &0.444     &0.452     &0.458 \\
4 &0.466     &0.473     &0.477     &0.479 \\
5 &0.489     &0.490     &0.491     &0.492 \\
6 &0.499     &0.499     &0.498     &0.498 \\
7 &0.503     &0.502     &0.501     &0.501 \\
8 &0.504     &0.503     &0.502     &0.502 \\ \hline
\end{tabular}

\

\

Table.3

\

\begin{tabular}{|c|c|c|c|c|}\hline
$B[T]$ &$\Delta[K]$ \\ \hline
5  &$-$0.017 \\
10 &0.142 \\
15 &0.285 \\
20 &0.398 \\ \hline
\end{tabular}
\end{document}